\documentclass[sn-mathphys-num]{sn-jnl}

\usepackage{graphicx}%
\usepackage{multirow}%
\usepackage{amsmath,amssymb,amsfonts}%
\usepackage{amsthm}%
\usepackage{amsmath}
\usepackage{amssymb}
\usepackage{mathrsfs}%
\usepackage[title]{appendix}%
\usepackage{xcolor}%
\usepackage{textcomp}%
\usepackage{manyfoot}%
\usepackage{booktabs}%
\usepackage{algorithm}%
\usepackage{algorithmicx}%
\usepackage{algpseudocode}%
\usepackage{listings}%
\usepackage{xcolor}
\usepackage{anyfontsize}
\usepackage{overpic} 
\usepackage{xargs} 
\usepackage{graphicx}
\usepackage{subcaption}
\usepackage{float}
\usepackage{url}

\begin{document}

\title[Article Title]{AI Foundation Model for Heliophysics: Applications, Design, and Implementation
}


\author*[1,2]{\fnm{Sujit} \sur{Roy}}\email{sujit.roy@nasa.gov, tsingh14@gsu.edu}\equalcont{These authors contributed equally to this work.}
\author*[5]{\fnm{Talwinder} \sur{Singh}}
\equalcont{These authors contributed equally to this work.}
\author[3]{\fnm{Marcus} \sur{Freitag}}\equalcont{These authors contributed equally to this work.}
\author[3]{\fnm{Johannes} \sur{Schmude}}\equalcont{These authors contributed equally to this work.}
\author[1]{\fnm{Rohit} \sur{Lal}}\equalcont{These authors contributed equally to this work.}
\author[6]{\fnm{Dinesha} \sur{Hegde}}\equalcont{These authors contributed equally to this work.}
\author[4]{\fnm{Soumya} \sur{Ranjan}}
\author[1]{\fnm{Amy} \sur{Lin}}
\author[1]{\fnm{Vishal} \sur{Gaur}}

\author[3]{\fnm{Etienne Eben} \sur{Vos}}
\author[1]{\fnm{Rinki} \sur{Ghosal}}

\author[8]{\fnm{Badri} \sur{Narayana Patro}}
\author[7]{\fnm{Berkay} \sur{Aydin}}
\author[6]{\fnm{Nikolai} \sur{Pogorelov}}
\author[3]{\fnm{Juan Bernabe} \sur{Moreno}}
\author[2]{\fnm{Manil} \sur{Maskey}}
\author[2]{\fnm{Rahul} \sur{Ramachandran}}

\affil[1]{\orgdiv{Earth System Science Center}, \orgname{The University of Alabama in Huntsville}, \city{Huntsville}, \state{Alabama}, \country{USA}}

\affil[2]{\orgname{NASA Marshall Space Flight Center},\city{Huntsville}, \state{Alabama}, \country{USA}}

\affil[3]{ \orgname{IBM Research}}
\affil[4]{ \orgname{Development Seed, Washington, DC, USA}}
\affil[5]{ \orgname{Department of Physics \& Astronomy, Georgia State University}}
\affil[6]{\orgname{Center for Space Plasma and Aeronomic Research (CSPAR), The University of Alabama in Huntsville, Huntsville, AL, USA}}
\affil[7]{\orgname{Department of Computer Science, Georgia State University, Atlanta, United States}}
\affil[8]{\orgname{Microsoft Research, India}}


\abstract{
Deep learning-based methods have been widely researched in the areas of language and vision, demonstrating their capacity to understand long sequences of data and their usefulness in numerous helio-physics applications. Foundation models (FMs), which are pre-trained on a large-scale datasets, form the basis for a variety of downstream tasks. These models, especially those based on transformers in vision and language, show exceptional potential for adapting to a wide range of downstream applications. In this paper, we provide our perspective on the criteria for designing an FM for heliophysics and associated challenges and applications using the Solar Dynamics Observatory (SDO) dataset. We believe that this is the first study to design an FM in the domain of heliophysics.
}




\maketitle

\section{Introduction}\label{sec1}

\begin{figure*}[!ht]
\centering
\begin{overpic}[width=0.9\textwidth]{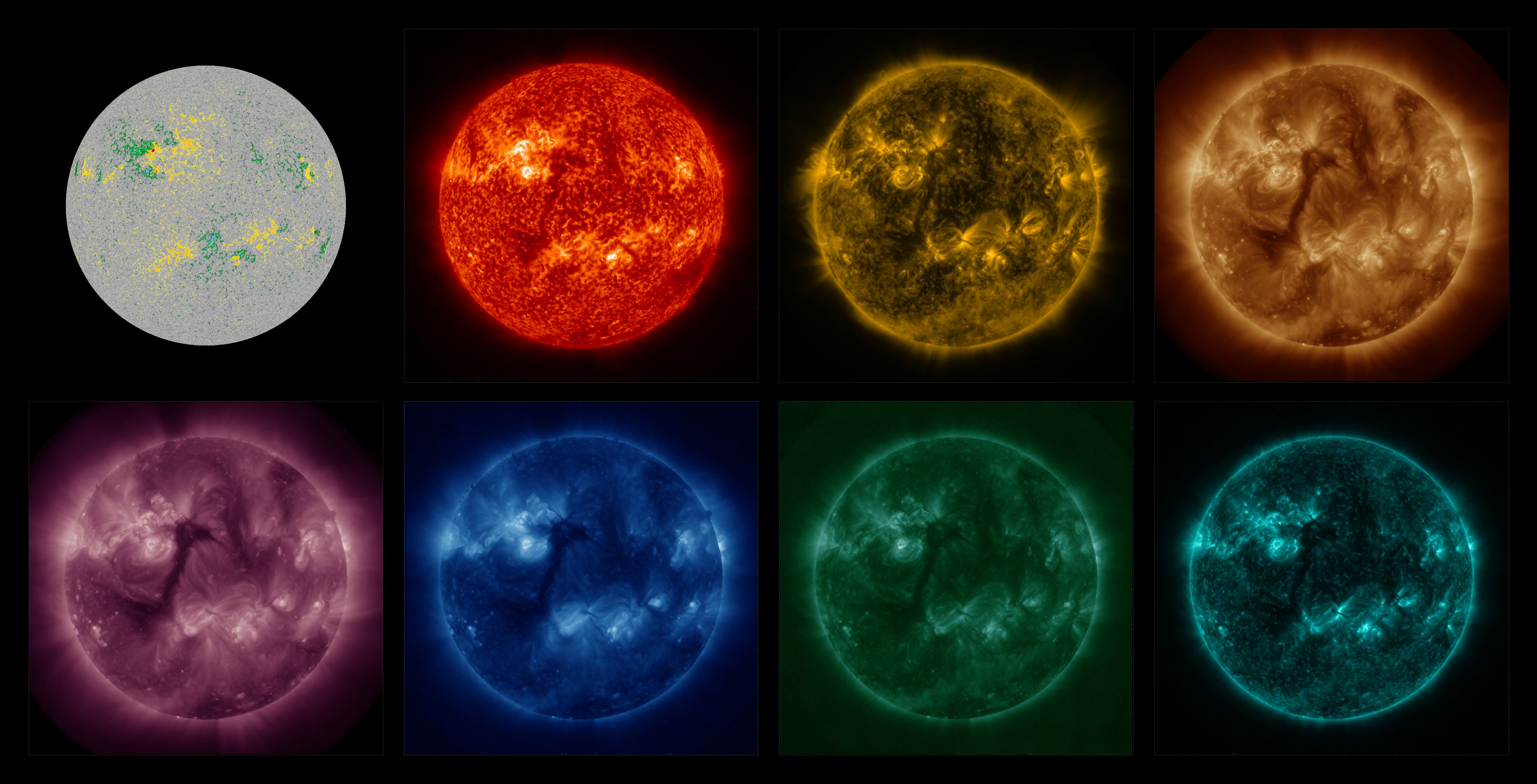} 
    \put(18,26){\color{white}\bfseries HMI}
    \put(43,26){\color{white}\bfseries 304}
    \put(68,26){\color{white}\bfseries 171}
    \put(93,26){\color{white}\bfseries 193}
    \put(18,1){\color{white}\bfseries 211}
    \put(43,1){\color{white}\bfseries 335}
    \put(68,1){\color{white}\bfseries 94}
    \put(93,1){\color{white}\bfseries 131}
\end{overpic}
\caption{Data captured by HMI and AIA instruments aboard SDO on 2012-01-30 at 22:12 UT. The top left panel shows the line-of-sight magnetogram from HMI and the rest of the panels show Sun in the seven EUV wavelengths from AIA. The wavelengths in $\textup{\AA}$  are mentioned in the bottom right corner of each panel.}
\label{fig:gooddata}
\end{figure*}
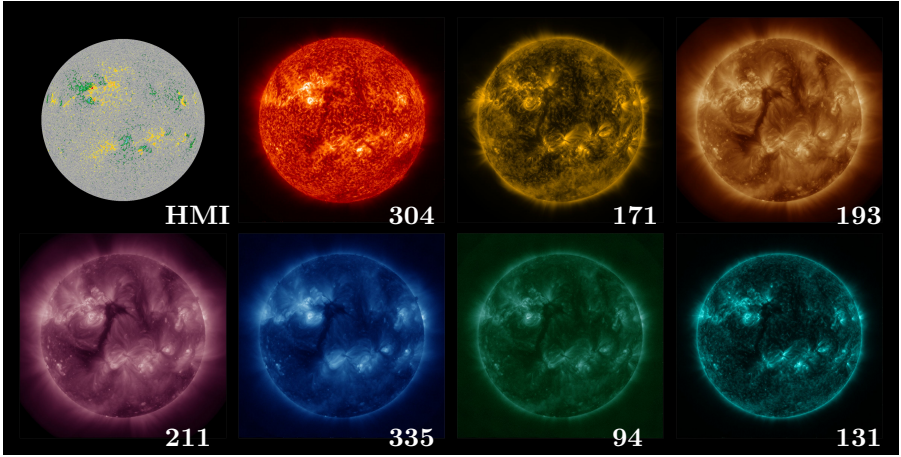

Since the launch of the Solar Dynamics Observatory (SDO), it has been providing a continuous high-resolution observation of the Sun \cite{Pesnell12}. SDO has been a very fruitful mission with $\approx$6000 peer-reviewed papers published using its data \citep{SDO_papers}. SDO's Helioseismic and Magnetic Imager \citep[HMI;][]{Schou12} provides high resolution (0.5 arcsec pixels) vector magnetograms and Doppler images, which are used to study the evolution of the magnetic field and plasma velocity on the solar surface respectively. The Atmospheric Imaging Assembly \citep[AIA;][]{Lemen12} onboard SDO provides full-disk high-resolution (0.6 arcsec pixels) images of the Sun in seven extreme ultraviolet (EUV) wavelengths that can capture plasma in the solar atmosphere starting from the chromosphere to the corona with temperatures from $\approx$20,000 K to more than 20 million K. AIA also images the Sun in two ultraviolet (UV) and one visible wavelength bands. AIA data have been instrumental in understanding the evolution of various coronal structures, such as coronal holes (CHs) \citep[][ and references therein]{Heinemann19}, active regions (ARs) \citep[][ and references therein]{Driel-Gesztelyi15}, coronal loops \citep[][ and references therein]{Reale14}, etc. This data has also led to better understanding of coronal processes such as EUV wave propagation \citep[][ and references therein]{Liu14}, solar flare eruptions \citep[][ and references therein]{Benz16}, filament eruptions \citep[][ and references therein]{Parenti14}, etc. Figure~\ref{fig:gooddata} shows the line-of-sight (LOS) magnetogram captured by HMI in the top-left panel. Other panels show the Sun observed in seven EUV wavelengths by AIA. Wavelengths in $\textup{\AA}$ are denoted in the bottom right corners of all panels.

SDO has produced (and continues to produce) an immense amount of data during its lifetime, sending down nearly 1.5 terabytes of data daily. SDO's large database provides a perfect opportunity to use machine learning (ML) based methods to solve many of the outstanding problems in solar and heliospheric physics. In this paper, we discuss how a class of ML models called Foundation Models (FMs) can be trained using the SDO data, detailing the implementation, planned downstream tasks, and preliminary results.

FMs generally consist of an encoder \( f_\theta \) and a decoder \( g_\phi \), where \( \theta \) and \( \phi \) represent the model parameters. Pre-training the FM involves optimizing
\[ g_\phi \circ f_\theta\]
using a self-supervised learning approach \citep{mukkavilli2023ai}. When fine-tuning for specific downstream tasks, the encoder \( f_\theta \) is pretained, but the decoder \( g_\phi \) is replaced by a task-specific decoder \( h_\psi \). However, if the fine tuning task involves very similar outcome like pretraining objective, we can keep the decoder and just change its head. In fine-tuning, the model then optimizes \( h_\psi \circ f_\theta \). Typically, decoders are shallow, whereas encoders can have millions or even billions of parameters, as they are designed to understand and represent the input data's complexity. This approach reduces the training burden by focusing on task-specific decoders. Furthermore, the pre-training data must be comprehensive enough that the learned distribution captures the variables and underlying principles necessary to generalize to various downstream tasks.

Recent studies \citep{climax, bodnar2024aurora, jakubik2023foundation, schmude2024prithviwxcfoundationmodel} have focused on developing a versatile AI system capable of handling various tasks by building FMs for earth sciences. Initially introduced in the domains of natural language processing (NLP) and computer vision (CV), FMs \citep{bommasani2021opportunities} are general-purpose AI models that have superseded specialized, task-specific models across numerous applications. Following the recent success of FMs in earth sciences, we propose an FM trained on SDO data, which will be geared towards solving problems in solar and heliospheric physics.  We believe that this is the first study to design an FM in the domain of heliophysics using high-resolution SDO data.

In Sec.~\ref{sec:applications}, we discuss the current state-of-the-art ML applications in solving problems in solar physics and heliophysics. There we also discuss why FMs can provide better solutions to these problems, along with the specific downstream tasks we have planned to implement with the pre-trained FM. In Sec.~\ref{sec:dataset}, we describe an ML-ready dataset we have prepared to train the FM. In Sec.~\ref{sec:methodology}, we detail the FM architecture, along with strategies and methods used in the training. In Sec.~\ref{sec:results}, we show the results we have achieved during the early training process. We provide our discussion and conclusions in Sec.~\ref{sec:conclusion}.

\begin{figure*}[!ht]
\centering
\includegraphics[width=0.9\textwidth]{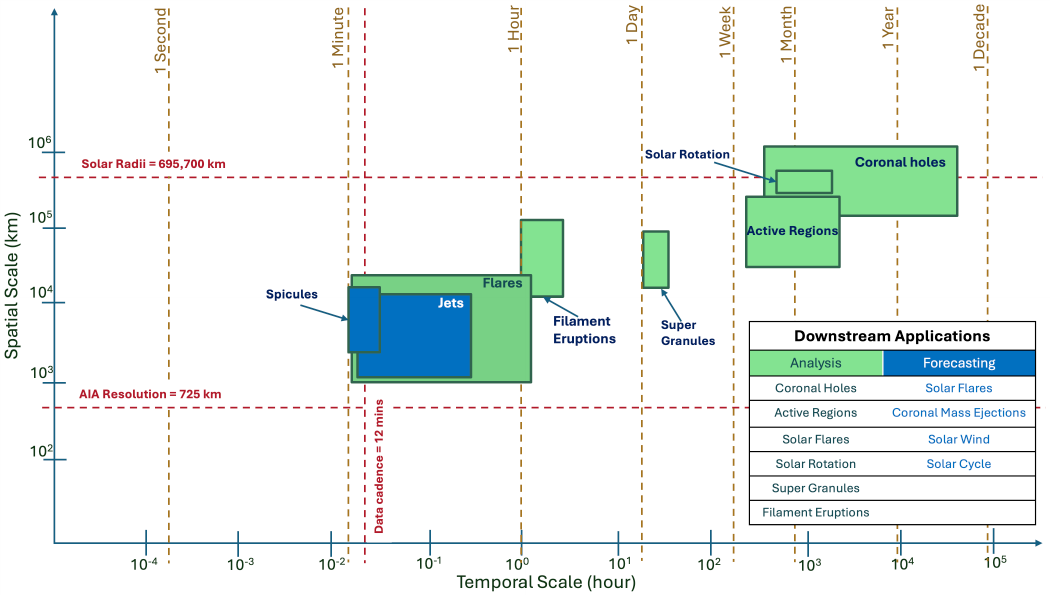} 
\caption{Visualization of selected downstream tasks with their spatial and temporal scales}
\label{fig:fmencdec}
\end{figure*}

\section{Applications \& State-of-the-Art Machine Learning Models}
\label{sec:applications}

In recent years, the SDO data has been used by several ML-based methods to achieve results superior to conventional methods. The application of ML methods using SDO data is very broad in the field of solar and heliospheric physics. Here, we list some of the applications.

ML methods have been very successful in feature detection on the Sun, be it CHs, ARs, prominences, or magnetic flux elements. For example, U-Net architectures have been successfully used to detect CHs and ARs in AIA full-disk images \citep{Illarionov18, Mackovjak21}. \citep{Jiang20} used similar architectures to identify and track solar magnetic flux elements observed in HMI magnetograms. Convolutional Neural Networks (CNNs) have been used successfully to delineate CH boundaries using a combination of AIA and HMI images \citep{Jarolim21}. CNN-based architectures have also been used to detect prominences and sunspots, in addition to CHs \citep{Baek21}. Deep Neural Networks (DNNs) have demonstrated 99.9\% accuracy in classifying solar phenomena such as quiet Sun, prominences, filaments, sunspots, and flare ribbons \citep{Armstrong19}. In addition, \citep{Inceogluetal:2022} has used an unsupervised ML method called pixel-wise k-means to detect CHs.

Forecasting solar flares is one of the areas heavily explored using ML methods trained on SDO data. Various studies have utilized non-potentiality proxies derived from LOS and vector magnetograms \citep{Ahmed13, Bobra15}, combining these with parameters from UV and EUV observations of ARs \citep{Nishizuka17, Nishizuka18}. Deep learning approaches have also been employed to forecast solar flares using magnetograms and intensitygrams without relying on external non-potentiality proxies of ARs \citep{Abed21}. Explainable models for flare prediction have been explored; for instance, \citep{Yi21} applied Gradient-weighted Class Activation Mapping (Grad-CAM) to identify influential regions on the solar surface that contribute to flare predictions, which were found to be regions close to polarity inversion lines.

The SDO data has also been used by ML methods to predict solar wind and space weather at Earth. For example, solar wind speed prediction at Earth has been addressed by training DNNs on sequences of AIA images \citep{Upendran20, Brown22}. In \citep{Rajuanddas:2021}, CNN based deep learning model has been used for solar wind forecasting. Additionally, \citep{Bernoux22} developed a DNN using SDO/AIA 193 \AA~images to obtain probabilistic forecasts of geomagnetic activity, specifically the Kp index.

ML methods, especially deep learning techniques, in combination with the SDO data have been used for advancements in 3D reconstruction of the solar corona. For example, \citep{Chifu21} employed a CNN to infer the $Z$-component of coronal loops solely from their two-dimensional shapes in EUV images. Meanwhile, \citep{Rahman23} demonstrated that Generative Adversarial Networks (GANs) can map photospheric magnetograms to electron density distributions at various atmospheric heights.  

Another category of applications that ML methods have made possible is the creation of synthetic data using image-to-image translation. For example, \citep{Szenicer19} trained a CNN with the SDO data to translate AIA images into disk-integrated irradiance data observed by SDO/EUV Variability Experiment (EVE; \citep{Woodsetal:2012}). The CNN-based architecture has also been used in autocalibration of SDO/AIA channels in \citep{DosSantos:2021}. In \citep{Kim19,Jeongetal:2020,Jeongetal:2022}, Conditional GANs (cGANs) have been used to generate synthetic LOS magnetograms from Solar TErrestrial RElations Observatory (STEREO)/EUVI \citep{howard2008} images. The same cGANs were used to generate He-I images from SDO/AIA \citep{Sonetal:2021}.  \citep{Salvatelli19,Salvatellietal:2022,Limetal:2021} explored the generation of synthetic EUV images in one wavelength from other wavelengths. In \citep{Galvezetal:2019, Parketal:2019} deep learning model has been used to translate SDO/HMI to SDO/AIA. Additionally, \citep{Felipe19} proposed a CNN-based model called \textit{FarNet}, which associates solar far-side maps obtained through helioseismic holography with solar magnetograms obtained when the far-side has rotated to face the Earth.

ML-based methods have also been used to improve the resolution of the existing data. For example, \citep{Diaz18} introduced \textit{Enhance}, a deep CNN that produces super-resolved continuum images and magnetograms for SDO/HMI, effectively increasing the per pixel resolution from $0.5''$ to $0.25''$. Similarly, \citep{Dou22} utilized GANs to improve the resolution of the Solar and Heliospheric Observatory (SOHO)/Michelson Doppler Imager (MDI;\citep{Scherreretal:1995}) magnetograms, to match the resolution of SDO/HMI magnetograms. 

Further, ML methods trained using SDO data have shown utility in solar cycle prediction \citep[e.g.][]{Okoh18, Benson20, Li21, Prasad22, Bizzarri22}. However, the results of these predictions vary widely and some of the predictions appear further from the truth than others, as the solar cycle 25 maxima approaches.

 
The applications listed above are not an exhaustive list of what is possible when ML methods are applied to SDO data. For a detailed review on this topic, see \cite{Asensioetal:2023}. The above mentioned applications convey that a broad range of solar and heliospheric problems can be targeted using the SDO data.  In Fig.~\ref{fig:fmencdec}, we show the temporal and spatial scales of various solar structures and processes. We notice that the majority of these can be studied with an SDO dataset with full spatial resolution and a time cadence of 12 minutes. An FM model trained with such a database can be fine-tuned for various downstream applications. These include the analysis of solar features such as CHs, ARs, and supergranules and processes such as solar flares, solar rotation, and filament eruptions. Besides analysis, an FM can be fine-tuned for forecasting specific downstream tasks like solar flare forecasting, filament eruption forecasting, and solar wind forecasting.

The FM approach envisioned in this work can be very beneficial to solve many of the outstanding problems in solar and heliospheric physics. Here, we list the potential benefits of using the FM approach:

\begin{enumerate}
    
    \item \textbf{Mitigation of Supervision Bottleneck:} By leveraging pre-training on extensive solar datasets, FMs reduce the reliance on labeled data. This mitigates the supervision bottleneck and improves real-world performance, particularly for tasks with forecasting of rare events like large solar flares. The scarcity of the data creates a number of issues for robust prediction (and even the data-driven detection) of solar weather events using traditional methods. The introduction of FMs have a great potential to solve this problem by removing the hefty early training process and have diverse application areas.
    
    \item \textbf{Versatility and Adaptability:} FMs can be easily fine-tuned for a variety of downstream tasks in solar and heliospheric physics, such as solar feature detection (coronal holes, sunspots, and active regions), space weather forecasting, and heliospheric modeling. The reduced need for labeled data and supervision makes these models adaptable to new datasets and tasks, providing significant value for researchers in the field.
    
    \item \textbf{Improved Generalization:} These models exhibit superior generalization capabilities, enabling them to handle data distribution shifts. This is especially useful for predicting rare or extreme events, such as solar flares, filament eruptions, and CMEs, where traditional models may struggle with generalizing from training data.
    
    \item \textbf{Scalability and Innovations:} FMs scale well with larger and higher-resolution datasets, making them ideal for handling multi-scale and multi-physics data from solar observations. This opens up new opportunities for high-resolution modeling of solar dynamics and heliospheric conditions, contributing to the next generation of data-driven models.

    \item \textbf{Potential for Integration with Multi-Modal Data:} The emerging trend of FMs presents opportunities for integrating multi-modal data from various solar and heliospheric observatories, such as combining SDO, SOHO, Parker Solar Probe, and other ground-based or space-based observations. This integration promises to broaden the scope and accuracy of future predictions and analyses of solar activity, space weather events, and heliospheric conditions by leveraging diverse datasets that include magnetograms, EUV images, and in-situ measurements.
    
\end{enumerate}

\section{Data and Preparation}
\label{sec:dataset}

In this work, we train the FM using time series data comprising solar images captured in seven EUV wavelengths (304 \AA, 171 \AA, 193 \AA, 211 \AA, 335 \AA, 94 \AA, and 131 \AA) by SDO/AIA, along with LOS and vector magnetograms obtained from SDO/HMI. Below, we describe how the SDO data was acquired, curated, and pre-processed to create an ML-ready dataset used for training the FM.

SDO data is publicly available through the Joint Science Operations Center (JSOC; \url{http://jsoc.stanford.edu}) as time series of various data products. The data products we have used to create our database are \textit{aia.lev1\_euv\_12s},  \textit{hmi.M\_720s}, and \textit{hmi.B\_720s}. The \textit{aia.lev1\_euv\_12s} series provides multi-wavelength AIA data with the resolution of 4096x4096 pixels and 12 second time cadence. The \textit{hmi.M\_720s} and \textit{hmi.B\_720s} series provide LOS and vector magnetograms, respectively, of the solar disk with the image resolution of 4096x4096 pixels and 12-minute cadence. To use HMI and AIA data together in the model, they should be spatially and temporally aligned. The following subsections describe the necessary pre-processing required to achieve this:

\begin{figure*}[!ht]
\centering
\includegraphics[width=\textwidth]{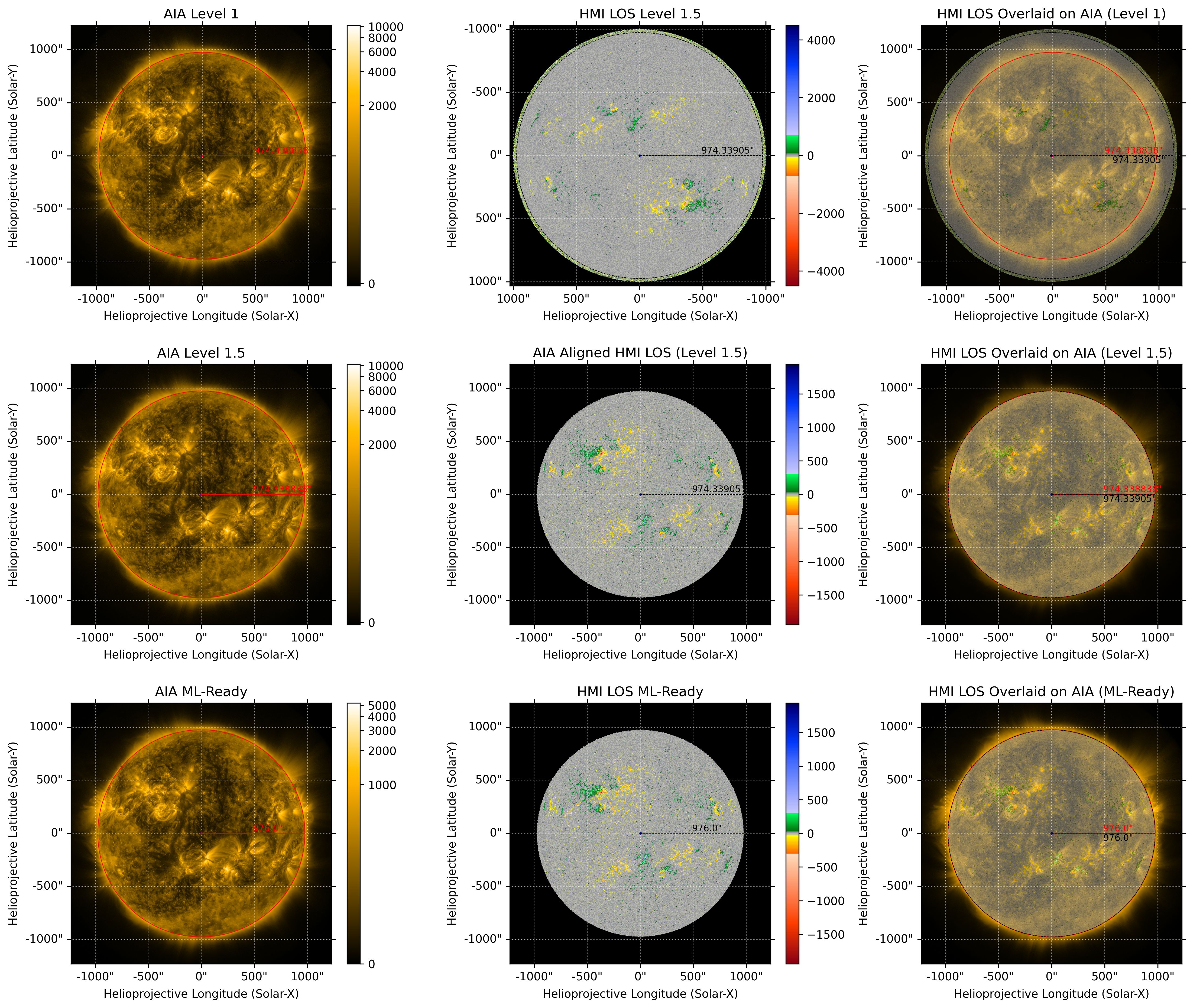} 
\caption{An example of the ML-ready data preparation steps for AIA 171 \AA\ and HMI LOS magnetogram on 2012-01-30 at 22:12 UT. Contours illustrate the image center, solar disk center, disk radius, and solar disk boundary. The top row shows the original AIA Level 1 image, HMI Level 1.5 magnetogram downloaded from JSOC, and HMI overlaid on AIA. The disk centers are misaligned with the image center (unregistered), and one dataset has a 180° roll, with noticeable plate scale differences. The middle row displays the registered AIA Level 1.5 image, HMI aligned with AIA, and HMI overlaid on AIA, showing corrected disk centers and plate scales. The bottom row presents the final ML-ready AIA and HMI images after exposure time normalization and orbital corrections for AIA, with the overlaid image showing proper alignment and a fixed disk radius of 976 arcsecs.}
\label{fig:prepsteps}
\end{figure*}

 \subsection{AIA}
 
The \textit{aia.lev1\_euv\_12s} series provided by JSOC contains level-1 data. This means that the images still include the roll angle of the satellite, i.e., the solar north-south axis is not aligned with the vertical y-axis,  and each channel may have a slightly different pixel scale. Hence it’s important to promote the AIA data from level 1 to level 1.5 to analyze the images from all the bands together. The promotion to level-1.5 involves updating the pointing keywords, removing the roll angle, scaling the image to a resolution of 0.6 arcsec per pixel, and translating the image such that the center of the Sun is located in the center of the image. Besides these steps, exposure time normalization is an extra but necessary step during the promotion because AIA measurements have heterogeneous exposure time ranging from 0.05s to 2.9 seconds. We show the conversion of AIA data from level 1 to level 1.5 in the two top left panels of Figure~\ref{fig:prepsteps} using an example of AIA 171 data.

\begin{figure}[h]
    \centering
    \begin{subfigure}[b]{0.47\textwidth}
        \centering
        \includegraphics[width=\textwidth]{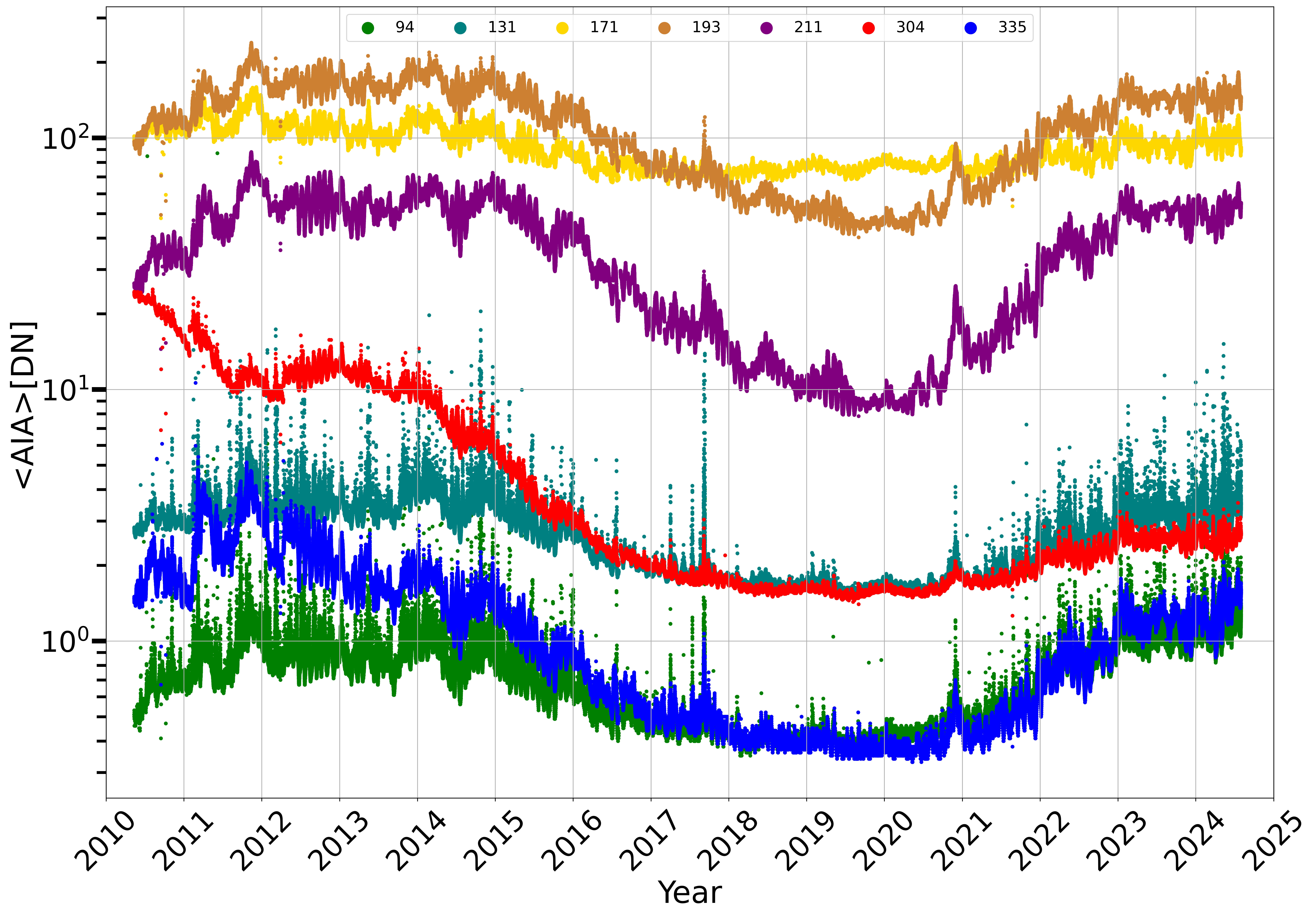}
    \end{subfigure}
    \hfill
    \begin{subfigure}[b]{0.47\textwidth}
        \centering
        \includegraphics[width=\textwidth]{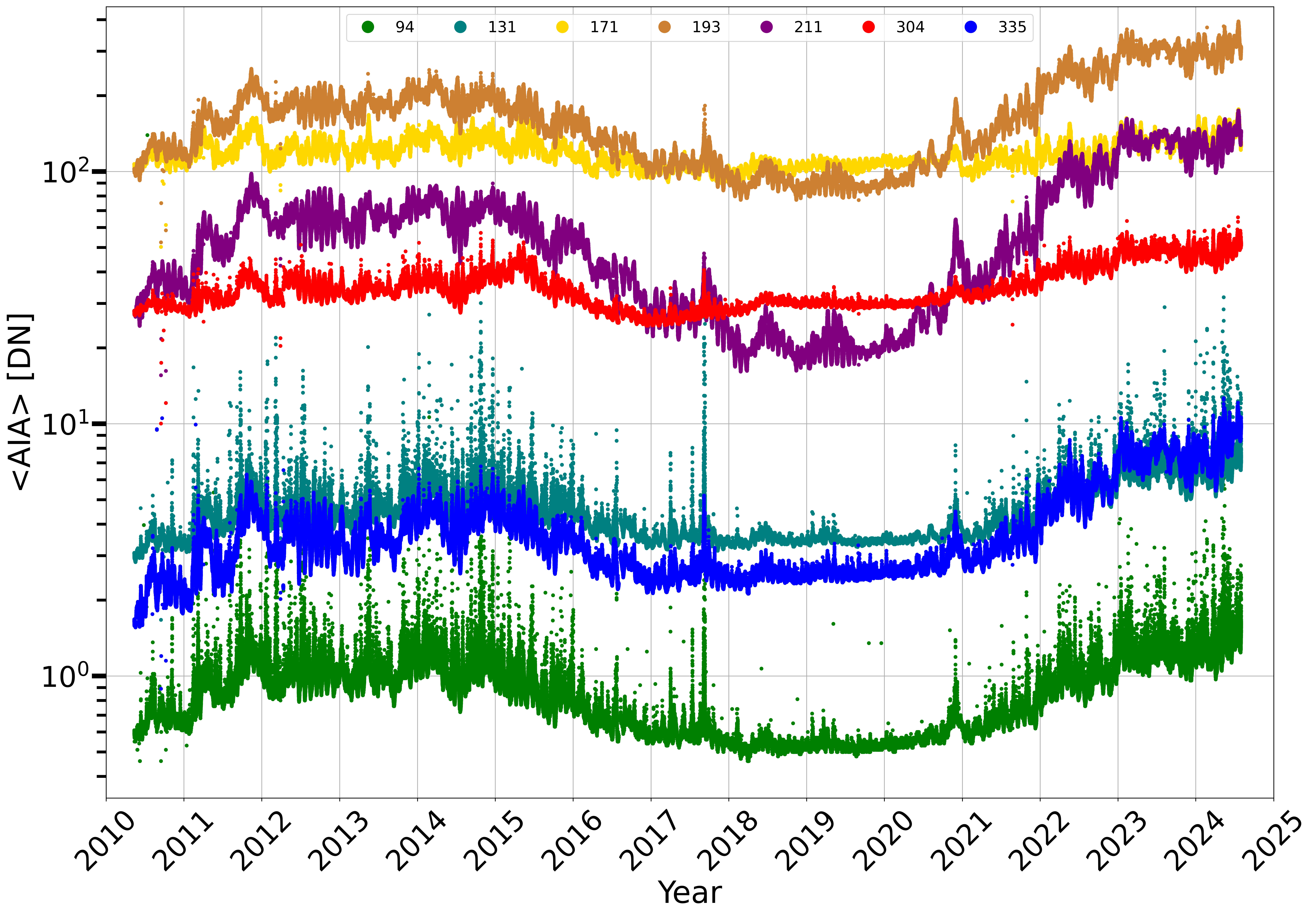}
    \end{subfigure}
    
    \caption{Mean of full-disk AIA images per second in seven wavelengths over time for before  (\textit{left panel}) and after (\textit{right panel}) the degradation correction.}
    \label{fig:degradation_correction}
\end{figure}

Since the database contains data throughout the SDO lifetime, CCD camera sensor degradation also needs to be taken into account. The table of correction parameters calculated by the AIA science team is made publicly available via JSOC. These parameters are a time series of scalars that can be multiplied by the full disk AIA data to rectify the instrument degradation. However, this method is problematic when the corrected values in some of the pixels exceed the instrument saturation value (16383 for AIA). We post-process the degradation-corrected image to make sure that none of the pixels reach a value larger than this limit. In Figure~\ref{fig:degradation_correction}, we plot the mean values of the full disk for each of the seven AIA wavelengths in level 1 (left panel) and level 1.5 with degradation correction (right panel) data. The variation of means over the years also reflects the solar cycle in which the high mean indicates strong solar activity. We notice that the degradation correction restores the higher activity in solar cycle 25 compared with the solar cycle 24.

One final step in making sure that the AIA data is ML-ready is to make the solar disk size the same in the whole database for all wavelengths, correcting for the elliptical orbit of the spacecraft. This step makes the solar disk radius as 976 arcsecs in all images. An example of this step is shown in the bottom-left panel of Figure~\ref{fig:prepsteps}.

 \subsection{HMI}
Though the observable HMI data provided from JSOC are level 1.5, the data have a different orientation due to the way the HMI is mounted on SDO. Therefore the HMI data needs to be re-projected to be spatially aligned with the level 1.5 AIA images. This re-projection is done using the \textit{reproject\_exact} function available in SunPy \citep{sunpy_community2020}. This function preserves the magnetic flux when spatially aligning the HMI magnetograms with the AIA images. An example of the re-projection is shown in the top two panels in the middle column of Figure~\ref{fig:prepsteps}. Finally, similar to AIA images, the HMI images also needed to be corrected for elliptical orbit variation and fix the solar disk size to 976 arcsecs throughout the database. This step is shown with an example in the bottom panel in the middle column of Figure~\ref{fig:prepsteps}. The panels in the right column of this image show that after our processing, the solar disk in the AIA and HMI images are well aligned.

\subsection{Temporal Alignment of HMI and AIA Data}
\begin{figure*}[!ht]
\centering
\includegraphics[width=0.9\textwidth]{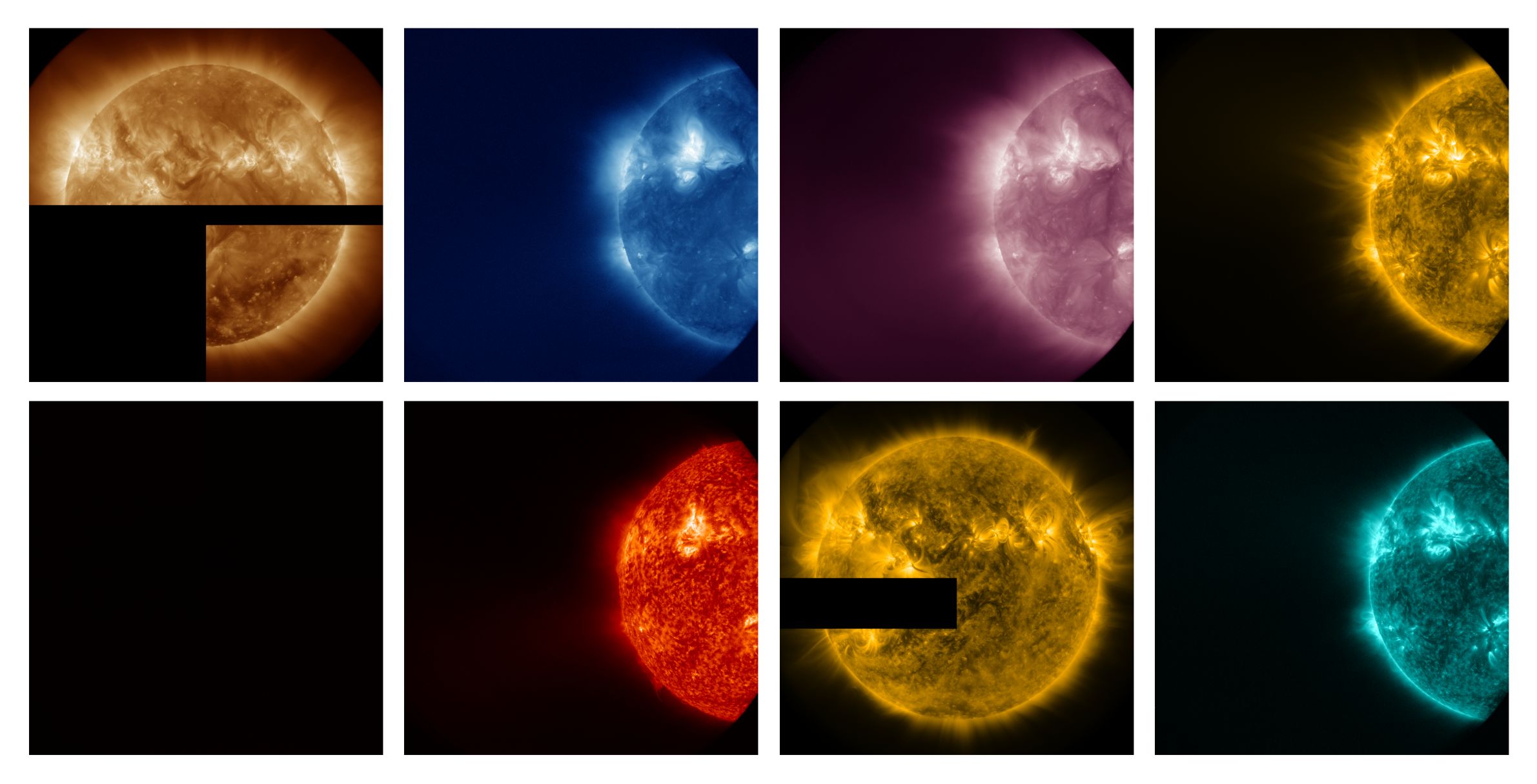} 
\caption{Bad AIA measurements due to a variety of reasons.}
\label{fig:baddata}
\end{figure*}

In our database, we have decided to keep the temporal resolution of the time series at 12 minutes. This is because the low-noise LOS and vector magnetograms are available with this cadence from HMI. Moreover, as discussed earlier, and shown in Fig.~\ref{fig:fmencdec}, our targeted downstream tasks for the FM can be satisfactorily implemented using a database with this cadence. For each of the timestamps in \textit{hmi.M\_720s} series, we find the corresponding AIA data at that time and add it to our database. If the quality of the AIA data for any of the seven wavelengths is not good at an exact timestamp of HMI, we search for a time within 2 minutes of the timestamp that contains good quality AIA data in all seven EUV wavelengths and include that data in our database. If we are unable to find good-quality AIA data within the two minutes range, we do not include the data in our database. We consider the quality of the AIA data to be good if the QUALITY flag in the file header is equal to zero. Non-zero values indicate different reasons for invalid data. In Figure~\ref{fig:baddata}, we give examples of when the AIA data was found to be of bad quality.

\subsection{Dataset Statistics}
\label{sec:Stats}
 

Our database contains ML-ready SDO data captured from May 13, 2010, to July 31, 2024. During this interval, there are about 2.9\% data unavailable (18,261 out of 623,280 total timestamps) for \textit{hmi.M\_720s} series. The processed level 1.5 AIA and HMI data are stored in hourly netCDF files in float32 format, with data shape of [5, 8, 4096, 4096]. Each netCDF file is about 2.2 GB, and the total size of the data for training is about 257 TB. 


\section{Methodology}
\label{sec:methodology}


\subsection{Sampling strategy}
A key consideration when pretraining an FM is to use a representative dataset that contains sufficient examples of both common and rare events.  This will enable the FM to efficiently learn characteristics and representations of all solar phenomena that could be relevant to downstream tasks.  What constitutes common and rare events on the Sun depends on factors like where we are in the solar cycle and what the focus of a particular downstream task is.  For example, solar minimum is characterized by periods of little solar activity, whereas solar maximum has ample activity in the form of ARs, CHs, flares, etc.  \citep{Svalgaard:2013}.  However, events such as large CMEs are relatively rare in comparison with other events.  The relevance of events, or regions on the Sun, also depends on the downstream application of interest, e.g. studies on magnetic flux would focus more on ARs and not consider granulation from quiet regions on the Sun \citep{MacTaggart:2021}, whereas studies of the solar wind would focus on various explosive events, but in particular on CHs \citep{Wang:2019}.  This highlights the need for a dataset that isn't biased towards any prevalent phenomenon and that accounts for spatial and temporal dynamics.

Despite ongoing efforts, determining the optimal composition of training data in computer vision remains elusive. Most current practices adjust the data mixture heuristically, often increasing the proportion of high-quality images or underrepresented classes without providing detailed criteria \citep{jakubik2023foundation}. Additionally, it's challenging to know whether these data strategies will be effective before completing the training process.

We consider the task of pretraining a Vision Transformer (ViT) model \citep{VIT} \( p_\theta \) for image modeling tasks such as masked image reconstruction or autoregressive prediction. The training data set is defined as \( \mathcal{D}_{\text{train}} = \left\{ \mathcal{D}_i \right\}_{i=1}^M \), comprising \( M \) visual domains.
The training dataset contains \( Z \) events (images), with each domain \( \mathcal{D}_i \) containing \( Z_i \) events such that:
\begin{equation}
Z = \sum_{i=1}^M Z_i
\end{equation}
We define \( z \) as the total number of events to be sampled from the dataset \( \mathcal{D}_{\text{train}} \) during training. Each domain \( \mathcal{D}_i \) will contribute \( z_i \) events to the sampled set, where:
\begin{equation}
z = \sum_{i=1}^M z_i
\end{equation}
The mixture proportions are defined as:
\begin{equation}
r_i = \frac{z_i}{z}
\end{equation}
These proportions determine the probability of sampling an image from domain \( \mathcal{D}_i \).
 We sample images \( \mathbf{x} \) from the entire dataset \( \mathcal{D}_{\text{train}} \) based on the mixture proportions \( r = [r_1, r_2, \dots, r_M] \). Each image \( \mathbf{x} \) is selected from domain \( \mathcal{D}_i \) with probability \( r_i \).

The Vision Transformer (ViT) model processes each image as a sequence of these patches. The loss function incorporates the sampling approach based on \( z \) events and is defined over the sampled images. For masked image modeling where a subset of patches \( \mathcal{M} \) is masked, the model aims to reconstruct the original pixel values of these masked patches. 

\begin{equation}
\mathcal{L}_\theta = \frac{1}{z} \sum_{i=1}^M \sum_{\mathbf{x} \in \mathcal{D}_i} \mathbb{I}_{\text{sampled}}(\mathbf{x}) \left[ -\sum_{j \in \mathcal{M}} \log P_\theta\left( x_j \mid \mathbf{x}_{\setminus \mathcal{M}} \right) \right]
\end{equation}

where \( \mathcal{M} \) is the set of indices for the masked patches in image \( \mathbf{x} \),  \( x_j \) is the pixel content of the masked patch at position \( j \), \( \mathbf{x}_{\setminus \mathcal{M}} \) represents the image with patches in \( \mathcal{M} \) masked out, \( P_\theta\left( x_j \mid \mathbf{x}_{\setminus \mathcal{M}} \right) \) is the model's probability of reconstructing the original patch \( x_j \) given the unmasked portions of the image and the loss is summed over the masked patches and averaged over \( z \).

In case of Autoregressive Modeling,

\begin{equation}
\mathcal{L}_\theta = \frac{1}{z} \sum_{i=1}^M \sum_{\mathbf{x} \in \mathcal{D}_i} \mathbb{I}_{\text{sampled}}(\mathbf{x}) \left[ -\sum_{j=1}^N \log P_\theta\left( x_j \mid x_{1, \ldots, j-1} \right) \right]
\end{equation}

where \( N = 65,\!536 \) is the total number of patches per image, \( x_j \) is the \( j \)-th patch in the sequence, \( x_{1, \ldots, j-1} \) are all preceding patches, \( P_\theta\left( x_j \mid x_{1, \ldots, j-1} \right) \) is the probability of predicting patch \( x_j \) given all previous patches and the loss is summed over all patches in the image and averaged over \( z \).




\subsection{Data Normalization}

When working with HMI and AIA data that has a wide dynamic range, direct use of raw values can pose challenges for visualization and learning in machine learning models. To address these challenges, we introduce the signum log transformation for the following key reasons:

\begin{enumerate}
\item  \textbf{Improving Dynamic Range for Visualization}: The raw data often spans a large range of values, making it difficult to visualize without setting arbitrary thresholds. The signum log transformation compresses this range, making the data more visually interpretable while retaining the structure of the original information.
\item  \textbf{Invertibility}: A crucial property of the signum log transformation is that it is invertible. This means that after transforming the data, we can retrieve the original values by applying the inverse transformation. This is essential in scenarios where we need to interpret results in the original scale, such as when producing final outputs or inverting processed data for model predictions.
\item  \textbf{Ease of Learning for Machine Learning Models}: Machine learning models tend to perform better when learning smaller values. Working with raw data that spans orders of magnitude (e.g., from \textit{$\mathcal{O}(10)$} to \textit{$\mathcal{O}(10,000)$}) can cause instability or slow convergence in the model. The signum log transformation compresses these large values into a smaller range, making it easier for models to learn meaningful patterns.
\item  \textbf{Preserving Sign Information}: The signum log transformation preserves the sign of the original data (positive or negative), which is critical when dealing with HMI data, as both positive and negative values carry polarity information.
\end{enumerate}

The signum log transformation is denoted by:
$$
\begin{aligned}
& f(x)=\operatorname{sgn}(x) \cdot \log (1+|x|) \\
& f(x)=\left\{\begin{array}{cl}
-\log (1-x) & ; x \leqslant 0 \\
\log (1+x) ; & x>0
\end{array}\right. \\
& f^{-1}(x)=\left\{\begin{aligned}
-e^{-x}+1 & ; x \leq 0 \\
e^x-1 & ; x>0
\end{aligned}\right. \\
& f^{-1}(x)=\operatorname{sgn}(x)\left[e^{|x|}-1\right]
\end{aligned}
$$

\subsection{Training objectives}

FMs are trained with task-agnostic pretext tasks. That is, training tasks that are intrinsic to the data and do not depend on any downstream use case yet that let the model learn a representation that can be applied to a large number of downstream tasks. For vision models, the canonical pretext tasks are masked reconstruction \citep{he2022masked} as well as contrastive learning \citep{chen2020simple}. In the former case, the model is given an input image with large parts missing and has then to reconstruct the entire image. The approach is easily implemented on transformer architectures. While generalization to higher dimensional settings such as videos is straightforward \citep{feichtenhofer2022masked}, this comes at a cost in token counts which can become problematic for standard ViT architectures. Contrastive approaches typically depend on randomized augmentations of the data. The pretext task is then to align embeddings of data irrespective of augmentation. This has been found to be particularly useful in the multi-modal setting.

Masked reconstruction and contrastive learning are widely used pretext tasks. While this is partially due to the powerful representations one obtains using them, another reason for their wide adoption is also that they easily apply to any image dataset. Yet in cases where the underlying dataset has more structure than just that of an image or data cube, it is possible to consider domain-specific pretext tasks. Ideally, these should mirror some properties of the domain to prompt the learning of more powerful and domain-specific representations. A nice example of this is the pretext task of AtmoDist \citep{hoffmann2023atmodist}. Here, the authors considered weather data from the ERA5 reanalysis. Their model was pretrained to classify the temporal difference between data from two different time stamps. If we stick to the weather and climate domain, \cite{schmude2024prithviwxcfoundationmodel} was pretrained to predict data at future or present times given incomplete knowledge about the present. This combines masking with forecasting albeit in a different way to the work by \cite{feichtenhofer2022masked} cited above. Moreover, it reflects some fundamental properties of weather data: That data is naturally sparse as it is the output of a complex observational system consisting of weather stations, satellites etc.~Yet at the same time the forecasting problem and the associated understanding of temporal dynamics is absolutely fundamental as well.

Beyond weather and climate data one can look at satellite imagery. The Prithvi model of \cite{jakubik2023foundation} was trained with spatial masking. However, it is straightforward to generalize the masking principle to the reconstruction of certain satellite bands.

With this in mind we can finally turn to solar data. Clearly dynamics play an important role so one would expect forecasting to be a powerful pretext task. At the same there is a strong interest in instrument emulation as shown by the work of \cite{dash2022high}. Which of course raises the possibility of using instrument to instrument translation as a non-temporal pretext task. Naturally one should also consider masking and contrastive learning. Masking is particularly useful if one aims for a model that can operate on different spatial domains. That is, if the domain of the data at inference time might change from the whole disk to a subset of the sun, pretraining with a masking objective ensures that the model is ``comfortable'' with missing data. This might be of particular interest in the context of heliophysics where instruments generally observe the part of the sun accessible via line of sight. All in all one expects that some combination of masking, forecasting and instrument-to-instrument translation can be used as a powerful pretext task to ensure that an FM for the sun learns a powerful representation of instrument data that can be applied to many use cases.

\subsection{Vision Transformers}
Recent studies on vision transformer-based architectures have introduced various strategies to implement attention mechanisms. Attention layers contribute to model accuracy by capturing long-range dependencies, handling diverse data types, accommodating various biases, and allowing for parallel processing. The strategies adopted in recent works for attention implementation can be broadly divided into two categories, namely, 3D self-attention and 2D cross-attention. The key advantage of these approaches is their focus on reducing the heavy computational demands typically associated with attention mechanisms in this data-rich field. Techniques like window-based attention, as employed in the Swin transformer \citep{liu2022video}, reduce the computational load by dividing feature maps into windows and applying self-attention within each window. This method has proven effective in both non-spherical and spherical scenarios, showing strong performance. Another approach leverages variable aggregation and cross-attention to simplify the attention complexity found in the basic Vision Transformer (ViT) architecture \citep{VIT}. We utilized these attention-based strategies to train our model. In the next few paragraphs, we will discuss details about the approaches taken by us and corresponding outcomes.

 Before delving into the details of these attention strategies followed by us, one must note that one of the approaches that has been utilized majorly for pretraining an FM is masking. In this paradigm, a certain portion of the input data is masked, and the model learns to recreate the masked image. Moreover, masking is done in terms of patches of size $H \times W$, where $H$ and $W$ are dimensions in pixels. The reason behind the use of masking is firstly because it lowers the memory load of the model on the GPU so that one can fit larger batch sizes and secondly, it helps the model to learn spatial information effectively. However, token size becomes a constraint eventually for such a large model. Even large vision transformer models like Prithvi WxC \citep{schmude2024prithviwxcfoundationmodel}, which has around 2.3 billion parameters, uses approximately 60,000 tokens (the input image dimension is 576x360 and each token size is 2x2). Similarly, if we try to do the same with SDO images with 8 or 10 channels with 4K resolution - we will end up with more than 8 million tokens. However, training such a large model directly is very difficult as per current hardware configurations. To counter these challenges and improve local and global attention \citep{swin_transformer} et al. came up with a shifted window mechanism, where we pay attention to tokens specified in a window.

\begin{figure}[!h]
\centering
\includegraphics[width=0.9\textwidth]{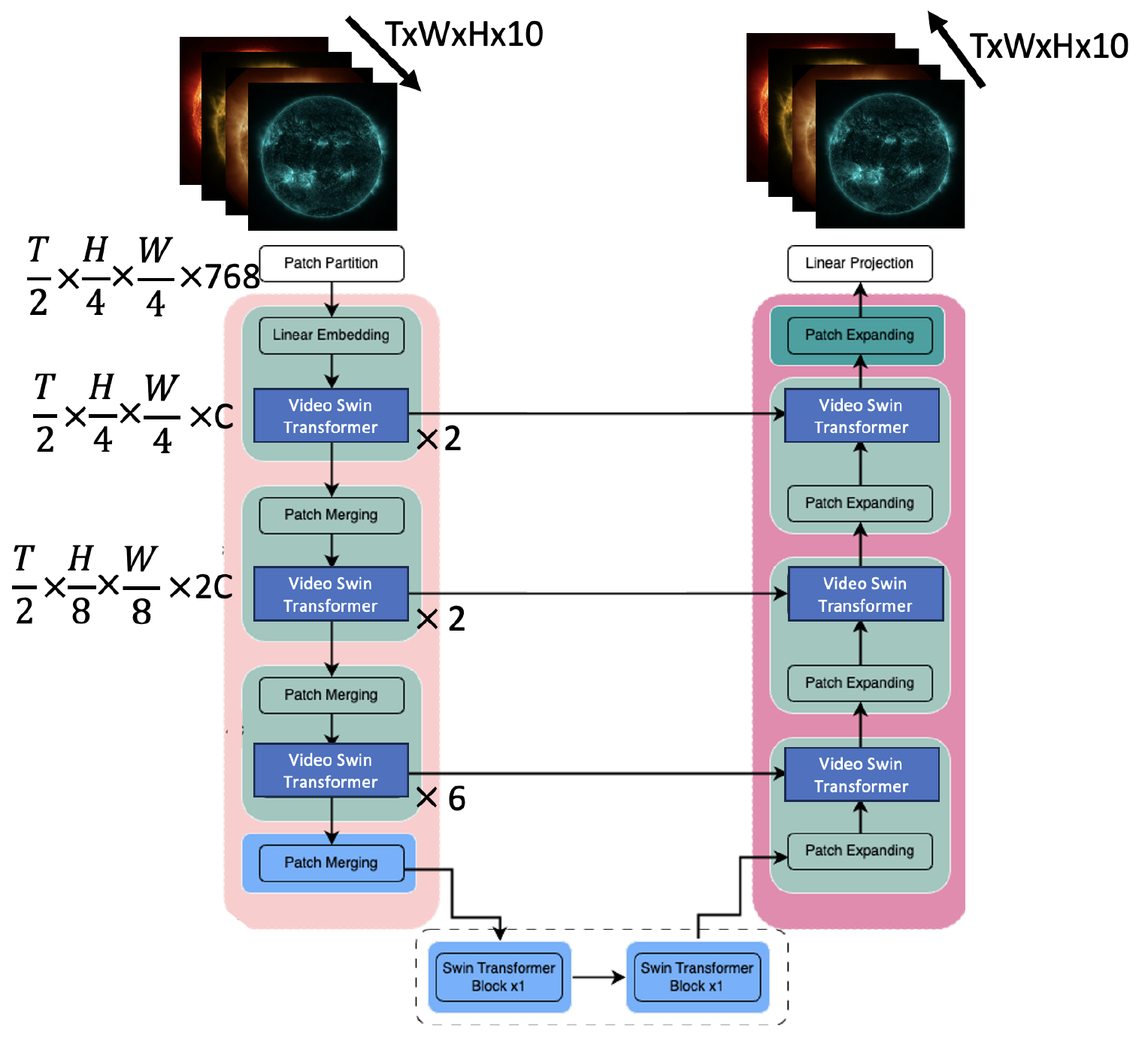} 
\caption{Architecture of $Model 1$, which uses video swin transformer blocks with U-Net alike structure for training on SDO data. The model takes two input time steps and forecast the next time step.}
\label{fig:swin}
\end{figure}

 Based on this logic we tried to train an architecture $Model 1$ inspired by \citep{cao2022swin} where we masked 50\% of the image. There were two objectives of the model- one to predict the mask and the second to generate an image given a lead time. Lead time is defined as $\Delta T$ where T can be between 1-5, i.e. 12-60 mins. Figure \ref{fig:swin} shows the architecture used to train the model is heavily inspired from \citep{swin_transformer}. The number of trainable parameters in this model is 1.8 billion, where the patch size is $16 \times 16 \times 2$. The model is given input of 2 timesteps. However, the experiments results were not good at all, with 30 epochs of training on 4 years of data. We also tried different masking strategies ranging from 10\% to 80\% at interval of 10. 

\begin{figure}[!h]
\centering
\includegraphics[width=0.9\textwidth]{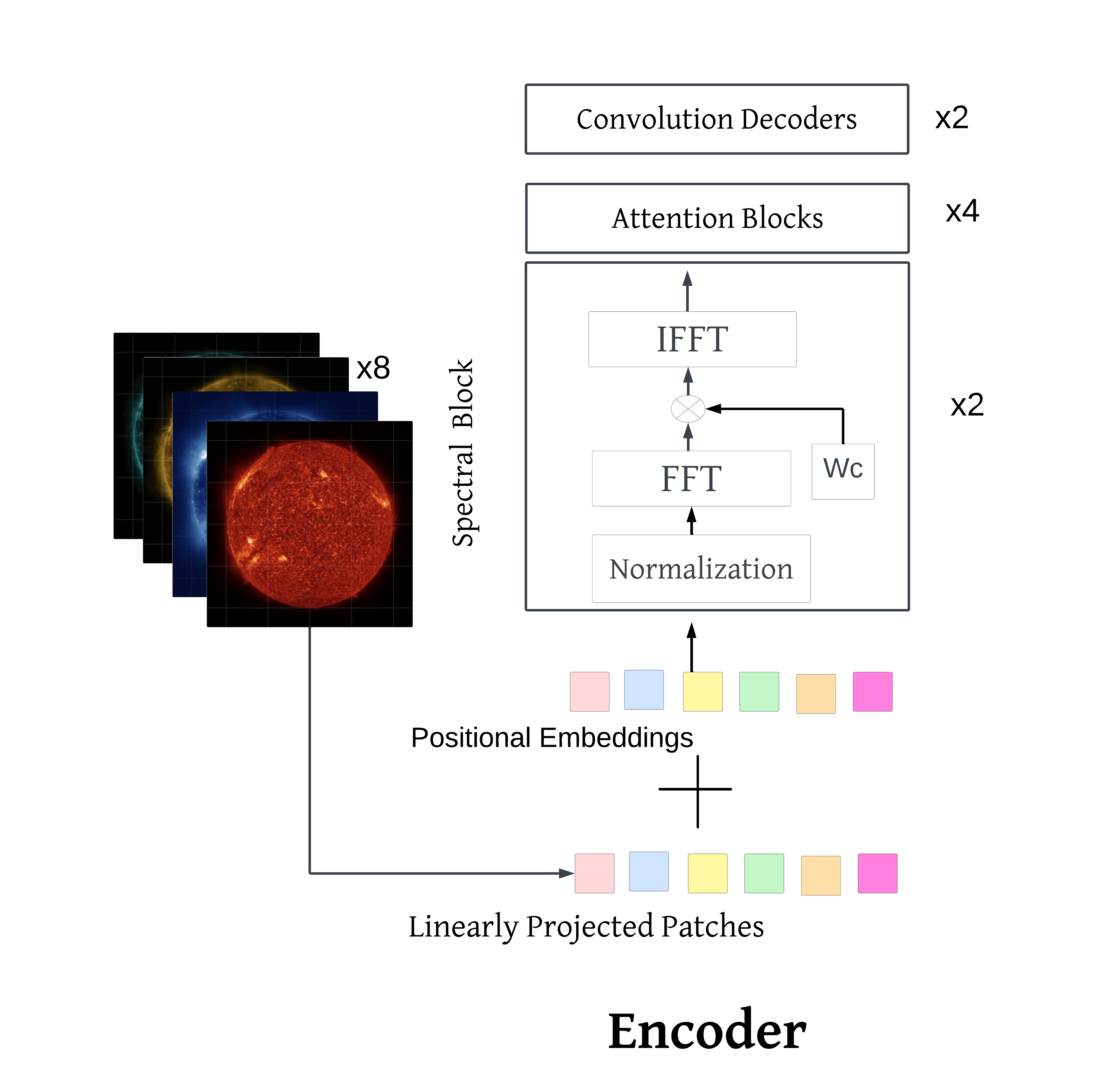} 
\caption{Architecture of spectral transformer as described in \citep{spectformer} used for training on SDO data. Convolution Blocks are added as decoders for the next step generation.}
\label{fig:spectralblock}
\end{figure}

However, the model trained with shifted windowed attention fails at generating the mask patches quickly. Hence, we tried another approach inspired by \citep{spectformer}. In this model, token mixing strategy is used by spectral blocks. This approach of token mixing makes the model faster. Using this methodology, post the patch embedding stage, the self-attention mechanism is replaced with a non-parametric Fourier transform, followed by a non-linear activation and a feed-forward network. This approach shows faster convergence to the model and better results for prediction for early epochs of training. The object of this model is to generate a lead time step only, i.e. forecasting model for time $t$. However, in this approach, masking technique can not be used as there will be missing information for Fast-Fourier Transforms ($FFT$). Figure \ref{fig:spectralblock} shows the basic architecture design of this model. Though the authors proposed an encoder-only model, a couple of convolution blocks are attached by us at the end to form decoder. We kept the convolution decoder shallow to enforce maximum learning for representation by encoders. In this case, the model is trained on $64 \times 64$ token size and the embedding space is of $1024$. However, it is previously mentioned that number of token going as low as $2 \times 2$ results in around 8 million tokens. To counter the issue, we adopted a hybrid architecture, explained in next section.

\subsubsection{Long-Short Spectral Transformer}
\begin{figure}[!h]
\centering
\includegraphics[width=0.9\textwidth]{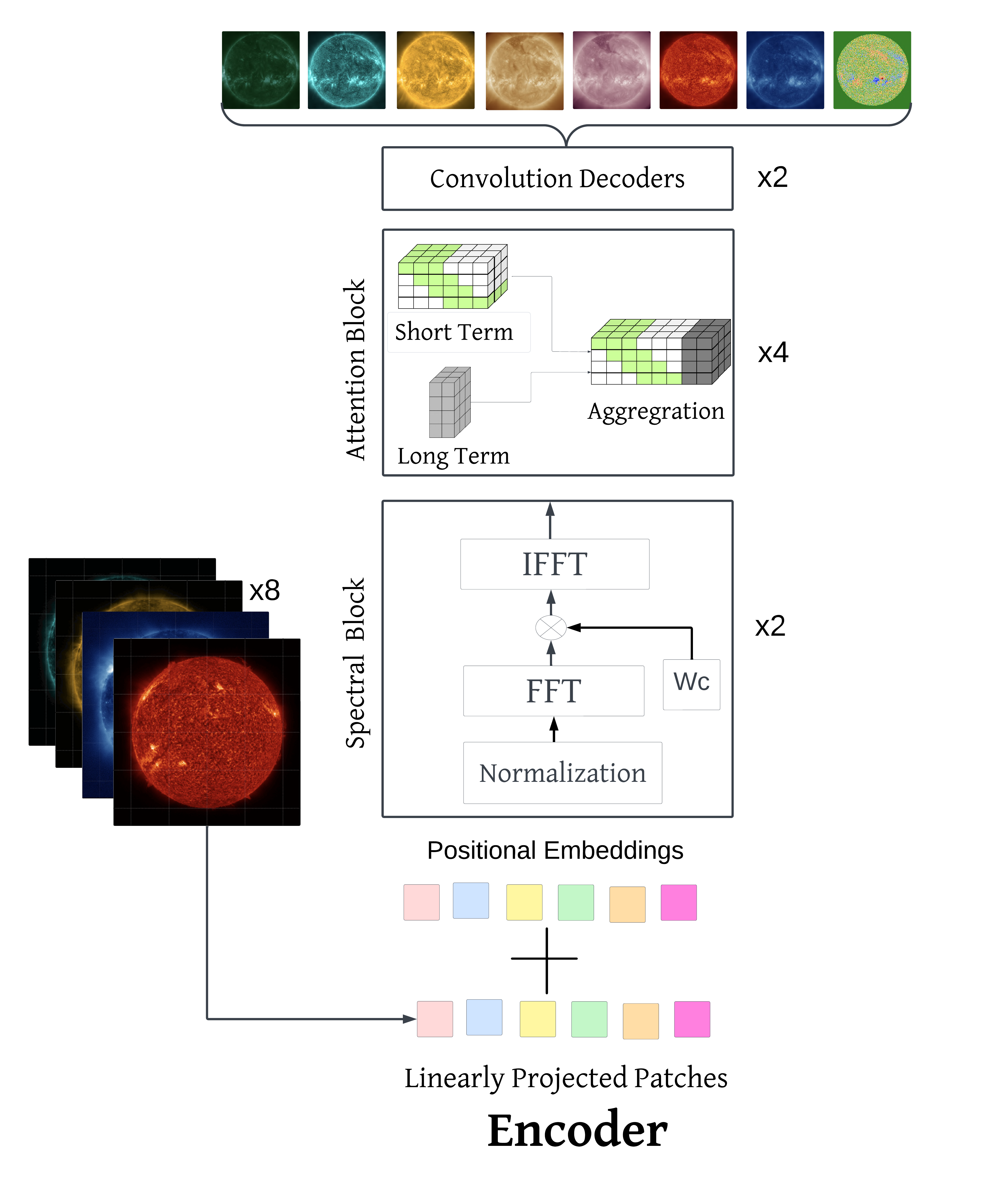} 
\caption{Architecture of Long-Short Spectral Transformer, where we are using a learnable weight parameter ($W_c$) after performing $FFT$ for estimating weights of frequency components. Post that we do inverse $FFT$ to bring the information to physical space. The attention head is designed on the principle of long-short attention, where we calculate short-term attention by sliding window, and long-range attention by dynamic projections}
\label{fig:lst_spect}
\end{figure}

To have larger number of sequence tokens, we came up with another strategy of mixing tokens and removing regular attention with the approach proposed by \citep{long_short_term_transformer}. In this approach, dynamic projection-based long-range attention is combined with local windowed short-term attention to effectively capture both global and local features in the input data. Specifically, long-range attention reduces the dimensionality of the Key and Value embeddings by projecting them into a smaller set of tokens using a projection matrix derived from the original Key embeddings. This mechanism allows the model to efficiently manage distant correlations across the sequence. On the other hand, short-term attention divides the input sequences into non-overlapping segments, focusing on capturing fine-grained local correlations. By merging these two mechanisms—long-range dynamic projection for distant dependencies and short-term attention for detailed local interactions—the model successfully addresses different scales of feature representation. By utilizing this approach we were able to train the model with a $16 \times 16$ patch size with a batch size of $4$ on A100 $80$ GB GPUs. Figure \ref{fig:lst_spect} shows the modified architecture of the model. In this version of model, we trained on embedding space of $1024$ along with 12 layers. In future, we will try to scale it to $8x8$ token size.



\subsection{Diffusion Models}

All the models discussed so far are largely deterministic in nature, i.e., given a pretext task such as masking, forecasting or some combination thereof one can train a transformer model on an RMSE objective or equivalent. While this approach is powerful and has led to many excellent results, it does have a few, known shortcomings.

To start, it is known that forecast models trained on an RMSE objective exhibit \emph{blurring} at longer lead times. That is, if one looks reasonably far into the future individual features start to lose focus. This should not be surprising. Intuitively speaking one can think of the space of all possible futures given an initial state. If this space of small futures depends on small changes in initial conditions, the model will have to effectively ``hedge its bets'' when making a prediction. It does so by predicting some averaged state between possible futures. Or to say this differently: In light of an uncertain future the optimal deterministic model trained on an RMSE (MAE) objective yields the average (median) of all possible futures.

A related issue is the ``double penalty'' of the RMSE objective. That is, if there are small, well defined features, a model trained on RMSE is heavily penalized if it misplaces such features by a few pixels: It is penalized both at the true location of the feature where it underpredicts as well as at the falsely predicted location where it overpredicts. Thus, in light of extreme values and uncertainty a model is incentivized to predict a smaller value over a larger area.

Both these issues can be addressed by moving to diffusion-based generative models \citep{sohl2015deep}. Roughly speaking, diffusion models train a denoising process that is capable of turning random noise into data samples. This denoising process can be conditioned on auxiliary data which allows us to model conditional distributions $p(x \vert y)$. The authors of \cite{price2023gencast, ruhling2024dyffusion} used this to train probabilistic forecast models $p(x_t \vert x_{t-1})$. Early applications for SDO data condition on the intensity of solar flares with the aim of generating synthetic solar flares \citep{ramunno2024solar}.

As the results of \cite{price2023gencast} show, in particular when compared with \cite{lam2022graphcast}, such model are much more capable of modeling fine structure and extremes in the data. In particular see \cite{lang2024ensembles} showing a diffusion model trained at 1 degree resolution exhibiting better fine structure than a deterministic model trained at 0.25 degrees resolution.

Diffusion models have some additional salient features. The authors of \cite{stanczukdiffusion} showed that diffusion models have implicit knowledge of the normal vector to the data manifold. This is of particular interest in the context of the question whether data-driven models accurately reflect the dynamics -- the physics -- of the underlying data. While conditioning on physically consistent data -- such as prior time steps or the measurements of other sensors -- massively reduces the risk of generating unphysical data, it is reassuring to know that such models have for the lack of a better word an aversion to generating samples away from the (training) data distribution. On a related note one should consider the work of \cite{lippe2024pde}. Here the authors used a diffusion process to improve the stability of neural PDE solvers during long rollouts. Essentially the diffusion model was trained to remove unphysical frequencies from AI-generated solutions to PDEs.

\subsection{Graph methods}

So far we have tacitly assumed that the data we are dealing with comes in the form of two-dimensional images or sequences thereof. This is a decent assumption since SDO data can easily be understood in image form and one can leverage results from the mainstream of computer vision architecture as well as the associated technological innovations and accelerations in deep learning toolkits. At the same time it is clear that the understanding of solar data in terms of images is limited. Large parts of the image model empty background space which is a waste of compute capacity. Moreover the structure of two-dimensional cartesian grids really does not respect the known structural priors of the data -- that we are modeling a spherical body for which we have data from line-of-sight instruments.

While transformers technically do not see gridded images but unordered sets of tokens, the known scaling issues of attention mechanisms used in transformers mean that such architectures often use optimizations that do tacitly assume that data lives on a rectangular grid. (See however \cite{schmude2024prithviwxcfoundationmodel}.) One alternative is to move to graph-based models. Here it is possible to define custom meshes and project the data on. This was done with great success in the work of \cite{lam2022graphcast}. In this paper the authors built a highly performant AI emulator for weather forecasting systems. At the heart of the emulator is an earth-encircling icosahedral hierarchical mesh. A graph neural network (GNN) then maps the data from a rectangular lat-lon grid onto this mesh, successively processes it through multiple layers and finally projects the data back onto the original grid.

While it is fairly straightforward to consider this application for solar physics -- either by projecting partial observations from instruments onto a spherical mesh or by having a 2d disc with changing data density from the center to the edges -- one should note that there are drawbacks. To start, the definition of such a mesh and network can be a considerable technical challenge. Moreover, GNNs while a very active research topic are not as much in the mainstream of AI as transformer architectures. Thus issues as their scaling performance are less understood.

\section{Results}\label{sec2}
\label{sec:results}

\begin{figure}[!ht]
\centering
\includegraphics[width=\textwidth]{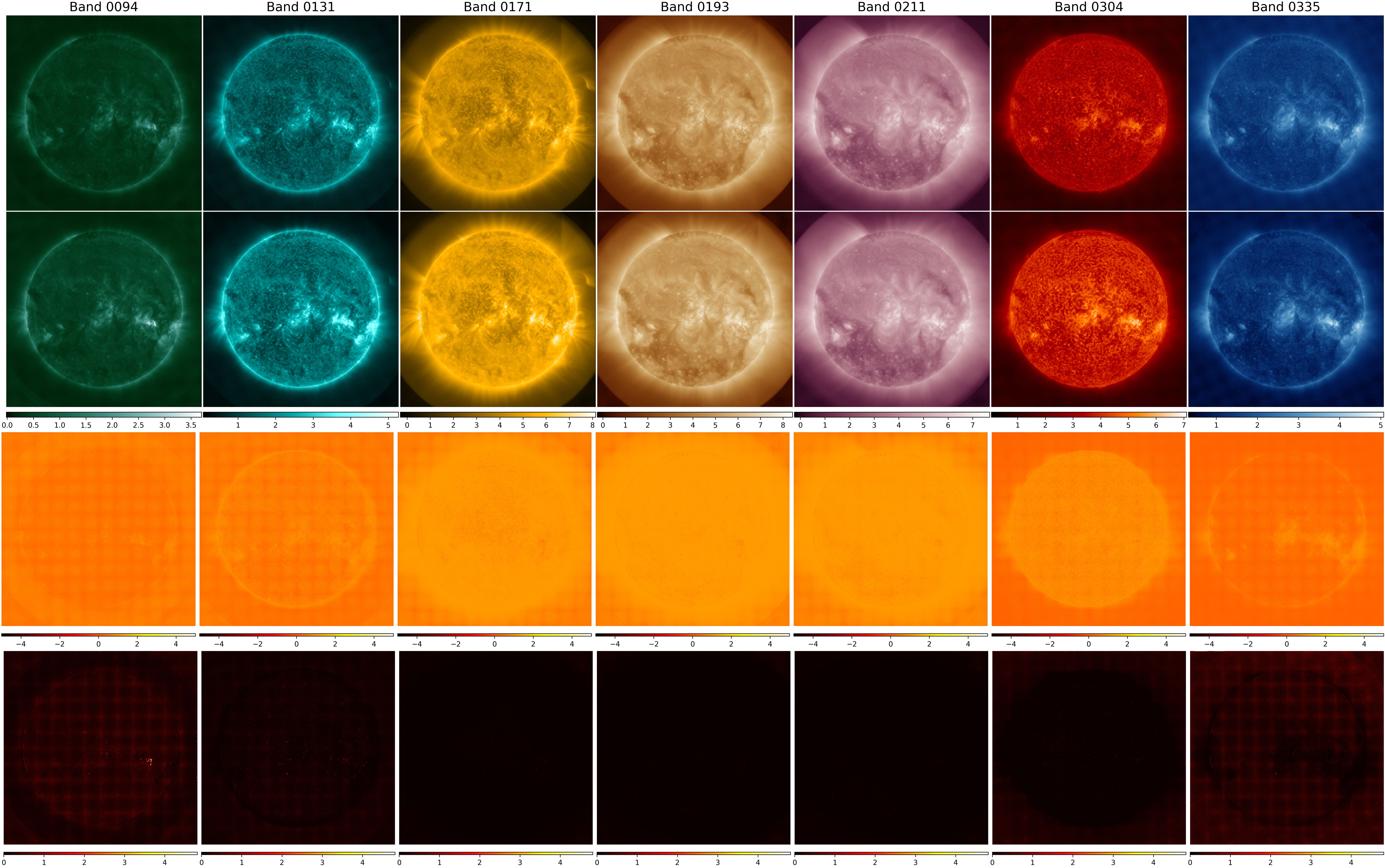} 
\caption{Forecast of AIA channels for next time steps, using 2 input timesteps. The first row is the ground truth for the model. The second row is the prediction from the model. The third row is the SSIM map between the ground truth and the prediction image. The fourth row is the squared error between the ground truth and the prediction image.}
\label{fig:resuls_lst}
\end{figure}
 
\begin{figure}[!ht]
\centering
\includegraphics[width=\textwidth]{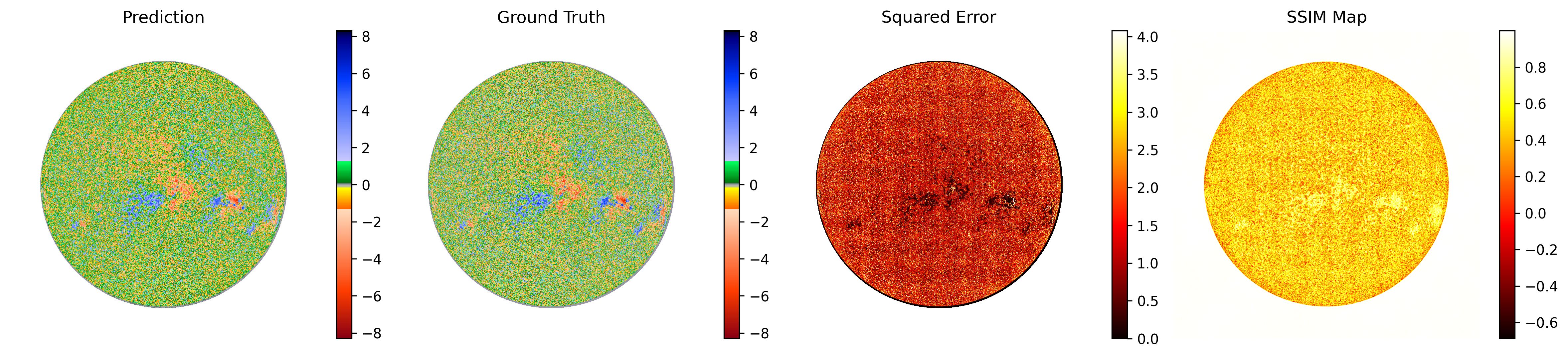} 
\caption{Forecast of HMI channels for next time steps. Ground truth denotes the image at time step $t$ in SDO data and prediction is the corresponding prediction from the model.  [From left] The first image is the prediction by the model, second is the ground truth. The third image is the squared difference between ground truth and prediction by the model and the fourth image is the SSIM map between ground truth and prediction.}
\label{fig:hmi_lst}
\end{figure}

\begin{figure}[!ht]
\centering
\includegraphics[width=\textwidth]{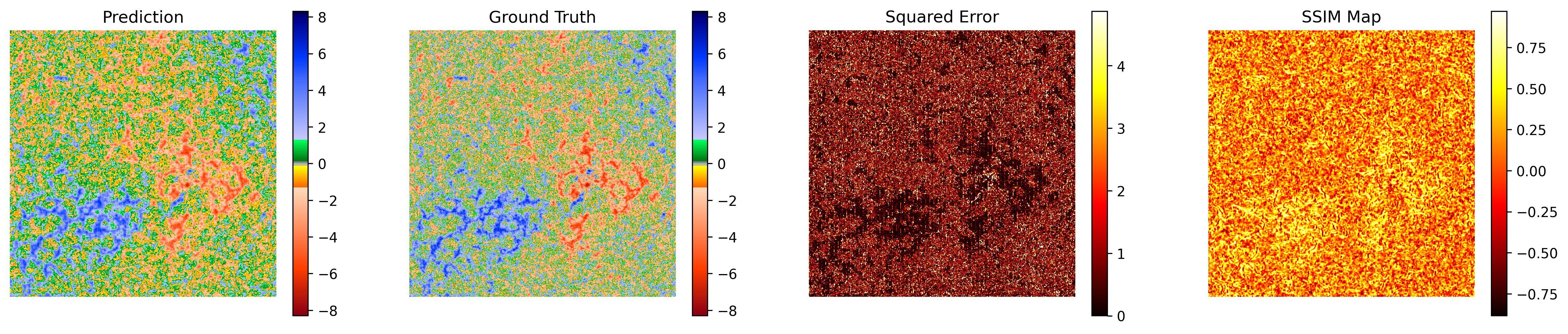} 
\caption{[From Left] Center crop of 1K resolution of (first image) prediction by the model, (second) ground truth. (third) squared difference between ground truth and prediction by the model and (fourth) SSIM map between ground truth and prediction.}
\label{fig:hmi_crop_lst}
\end{figure}

\subsection*{Initial results with Long-Short Spectral Transformer}
\textbf{Design of experiment:} To calculate initial results and the capability of the model to learn quickly, we designed experiments that would run for a maximum of 24-36 hours with a maximum of 16 nodes of GPU resources where each node had 4 A100 80GB GPU. For the results below the model was trained on 4 years of SDO data (2011-2015) and validation was performed on 6 months of data. Data was normalized by using signum log as described in section \ref{sec:Stats}.\\

Figure \ref{fig:resuls_lst} shows the ground truth, prediction, structural similarity index measures (SSIM) map, and the squared difference of both by using our modified long-short Spectral transformer \ref{fig:lst_spect}. The squared difference between the image is defined as $X = (Y-\hat{Y})^2$. The SSIM value for band 0171 is 0.83, for band 0193 is 0.90, and for band 0211 is 0.86. For other bands, the SSIM was between 0.4 to 0.65. The root mean squared error (RMSE) for band 0171 is 0.11, for band 0193 is 0.095, and for band 0211 is 0.10. With just 20 epochs of training on 4 years of data, the model shows remarkable performance. It is to be noted that, further training will be able to refine the image structure and also remove very light patchy artifacts from the model, which is primarily influenced by token size.

Figure \ref{fig:hmi_lst} shows the prediction of the HMI channel by using the same model, where the first image from left is the prediction, followed by ground truth, squared error, and SSIM map.  The bright patches in the SSIM map and the dark patches in the squared error map show that the model was able to reproduce the ARs with higher magnetic field strengths compared to the background quite Sun. To look at a finer scale, we performed a center crop of 1K resolution on the HMI band and performed the same comparison as shown in figure \ref{fig:hmi_crop_lst} and the images follow the same order as in \ref{fig:hmi_lst}. The SSIM score for the predicted image with respect to ground truth is 0.73 and the RMSE is 0.65. Since the quiet Sun contains scale, short-lived, dynamic magnetic flux elements, capturing the temporal evolution of these regions is relatively challenging than the large-scale ARs. However, further improvements in the FM architecture, training strategies, and inclusion of multiple optimization functions would improve the results.

\subsection*{Initial Results with Diffusion} 

\begin{figure}[!ht]
\centering
\includegraphics[width=0.99\textwidth]{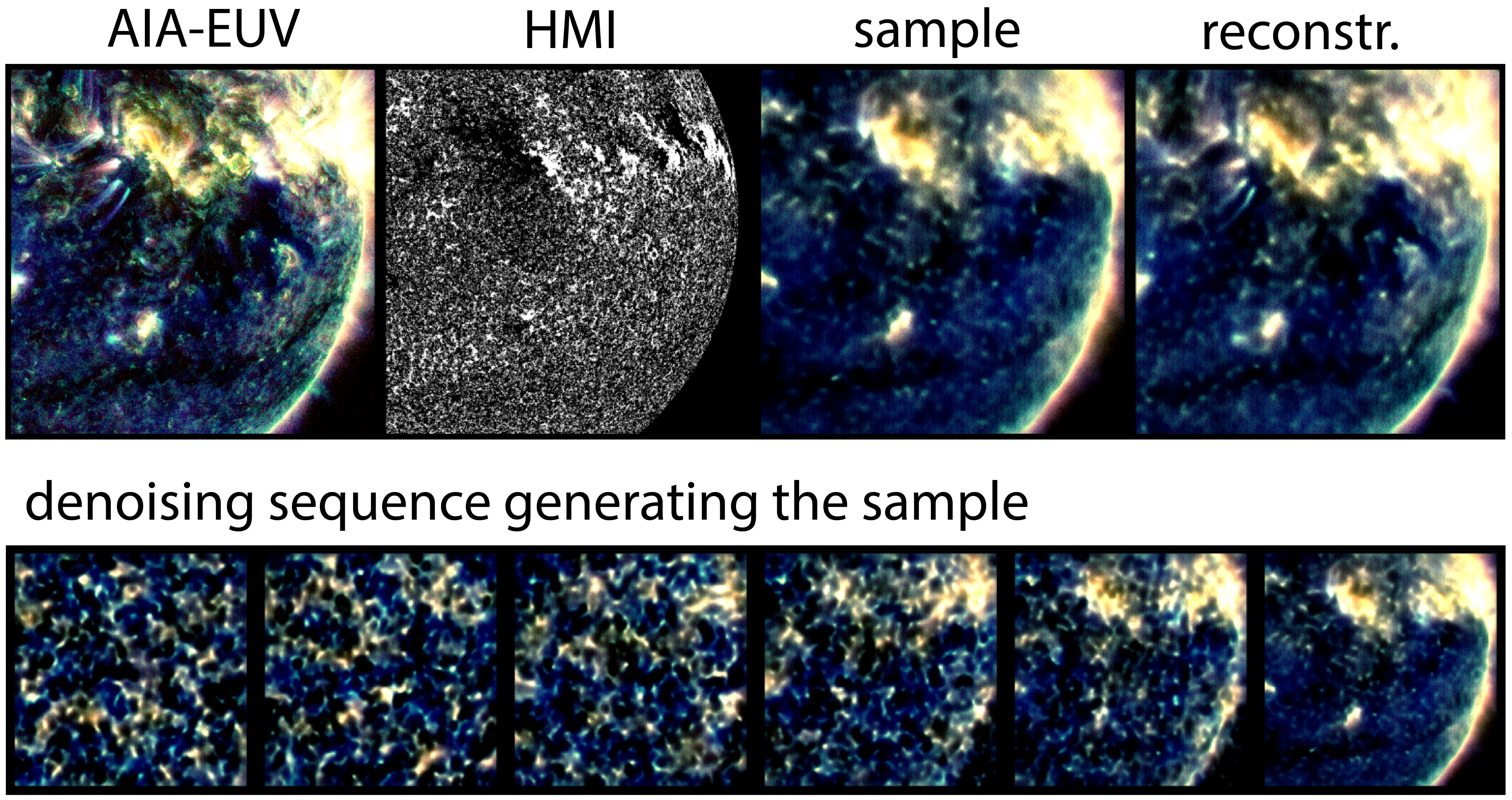} 
\caption{Generation of AIA-EUV channels through Latent Diffusion conditioned on HMI. Top row: 3-channel AIA-EUV image (channels 094, 131, and 171 displayed as RGB), corresponding HMI image used for conditioning the latent diffusion model, denoised sample generated from Gaussian noise, and reconstructed autoencoder AIA-EUV image. AIA-EUV and reconstructed images are shown for comparison only (other instances of these are used during training). Bottom row: Denoising sequence (7 timesteps) showing how the sample emerges from noise through successive denoising operations (1000 timesteps) conditioned on HMI.}
\label{fig:ldm_conditioning}
\end{figure}

In our current approach to diffusion models, we follow R. Rombach et al. \cite{rombach2022high} and first train shallow autoencoders to reduce image sizes before they are fed to the diffusion model. This latent diffusion model approach has been shown to “reach a near-optimal point between complexity reduction and detail preservation”, \cite{rombach2022high} and will allow us to eventually scale the model to the native $4096 \times 4096$ pixel resolution of SDO data. Initial experiments are done on downscaled versions and/or random crops of 3-channel SDO data (AIA-EUV channels 094, 131, and 171).

Among the many possibilities for conditioning the latent diffusion model, such as through class labels (e.g. strength of flares), semantic maps (e.g. position of active regions), or image-based conditioning (e.g. prior SDO timesteps, different SDO channels or different instruments), we chose one of the more challenging tasks: translation from HMI to AIA-EUV. HMI images are auto-encoded like the 3-channel SDO data or simply downscaled to the latent space dimensions through bilinear interpolation. We tried conditioning using either cross-attention or concatenation. Better results were achieved using concatenation.

Fig. \ref{fig:ldm_conditioning} shows an example of AIA-EUV generation conditioned on HMI. Here we used a 4-level autoencoder to encode AIA-EUV images of $512 \times 512 \times 3$ pixels to $32 \times 32 \times3$ pixels in latent space, where the latent diffusion model is applied. The HMI images were downscaled through bilinear interpolation to $32 \times 32$ pixels. Conditioning on HMI was achieved through concatenation.

\section{Discussion \& Conclusion}\label{sec12}
\label{sec:conclusion}

We presented the first study to design an FM in the domain of heliophysics. The development of an FM for heliophysics represents a significant step in leveraging deep learning techniques to advance our understanding of solar phenomena. One of the significant advantages of using an FM in heliophysics is the mitigation of the supervision bottleneck. Traditional supervised learning methods require extensive labeled datasets, which are challenging to obtain in heliophysics due to the rarity and complexity of certain solar events. By employing self-supervised learning during pre-training, the FM learns intrinsic features from the vast unlabeled SDO dataset, reducing the dependency on labeled data for downstream applications. This approach not only enhances the model's adaptability but also allows for better generalization across different types of solar phenomena. In this paper, we have tried to draw a scope for designing an FM using the SDO dataset. We have also shown preliminary experimental results by pre-training an encoder-decoder architecture on four years of high-resolution data from the Solar Dynamics Observatory (SDO). We have aimed to capture the intricate temporal and spatial patterns inherent in solar activities. The inclusion of eight bands — seven EUV channels from the Atmospheric Imaging Assembly (AIA) and one magnetogram from the Helioseismic and Magnetic Imager (HMI) — provides a comprehensive dataset that encapsulates various information of the solar atmosphere.

Considering state-of-the-art methods like Swin transformer \citep{liu2022video}, Vision Transformer \citep{VIT}, and spectral transformer \citep{spectformer}  and the challenges associated with them, we proposed a new approach for model design. keeping note of image size and data volume and limitations for larger token size and lower dimensions,  we also presented our approach that combines frequency-based information with long-short attention mechanisms, enabling the model to handle the dynamic nature of solar events. The integration of frequency-based components allows the model to capture local features, as well as global features for long-range dependencies. Additionally, by adding a long-short attention mechanism, we intend to maximize the visibility of more tokens by the model.

Despite the experimental constraints to find an optimized model, initial results are promising, demonstrating the model's capability to predict solar dynamics and its potential applicability to various downstream tasks. For instance, the FM shows proficiency in capturing the evolution of active regions and predicting solar flares by analyzing patterns in magnetic field configurations and coronal activities. This ability is critical for space weather forecasting, where accurate predictions of solar events can mitigate the impacts on satellite operations, communication systems, and power grids on Earth.

In conclusion, the development of an AI FM for heliophysics holds significant promise for advancing solar research and improving space weather forecasting. While challenges related to data volume, computational resources, and variability exist, the initial results are encouraging. By addressing these challenges through efficient data handling, robust training strategies, and comprehensive validation, the FM can become a versatile tool for the heliophysics community. Ongoing efforts will focus on refining the model, expanding its applications, and ultimately enhancing our ability to predict and understand complex solar phenomena that impact our technological society.
\backmatter


\bmhead{Acknowledgements}
The Authors acknowledge the National Artificial Intelligence Research Resource (NAIRR) Pilot and NVIDIA for providing support under grant no. NAIRR240178. The authors would also like to thank Jülich Supercomputing Centre (JSC) and  NASA Advanced Supercomputing (NAS) Division for their compute support.

\bibliography{sn-bibliography}


\end{document}